\documentstyle[11pt,gh2001-asp,twoside,epsf]{article}
\def\om{\Omega_p}
\def\len{a_B}
\def\lag{D_L}
\def\vpd{{\cal R}}
\def\kin{{\cal V}}
\def\pin{{\cal X}}
\def\mkin{{\cal K}}
\def\mpin{{\cal P}}
\def\msmt{{\cal S}}
\def\DV{{\Delta V}}

\def\spose#1{\hbox to 0pt{#1\hss}}
\def\gtsim{\mathrel{\spose{\lower.5ex \hbox{$\mathchar"218$}}
     \raise.4ex\hbox{$\mathchar"13E$}}}
\def\ltsim{\mathrel{\spose{\lower.5ex\hbox{$\mathchar"218$}}
     \raise.4ex\hbox{$\mathchar"13C$}}}

\def\degrees{^\circ}

\def\arcsec{^{\prime\prime}}
\def\kms{$\mathrm km\;s^{-1}$}
\def\kmsa{$\mathrm km\;s^{-1}\,arcsec^{-1}$}
\def\kmsk{$\mathrm {km}~\mathrm{s}^{-1}~ \mathrm{kpc}^{-1}$}

\def\edcomment#1{\iffalse\marginpar{\raggedright\sl#1\/}\else\relax\fi}
\reversemarginpar

\begin{document}
\title{Bar Dynamical Friction and Disk Galaxy Dark Matter Content}
\author{Victor P. Debattista}
\affil{Astronomisches Institut, 
     Universit\"at Basel, 
     Venusstrasse 7,
     CH-4102 Binningen, Switzerland}

\begin{abstract}
I present some new results, from simulations and observations, of the
constraints bar pattern speeds place on the dark matter content of
disk galaxies.
\end{abstract}

\section{Introduction}
Near infra-red images reveal bars in over $50\%$ of disk galaxies 
(Knapen et al. 2000; Eskridge et al. 2000).  The principal 
dynamical quantity for barred (SB) galaxies is the pattern speed of the bar, 
$\om$, usually parametrized by the ratio $\vpd \equiv \lag/\len$ (where 
$\lag$ is the corotation radius and $\len$ is the semi-major axis of the bar).
A bar is termed fast when $1.0 \leq \vpd \ltsim 1.4$, while, for a larger 
value of $\vpd$, a bar is said to be slow.  All measurements of bar pattern 
speeds have found fast bars (e.g. Gerssen 2002), a result which has 
been interpreted as evidence for maximum disks in SB galaxies (Debattista \&
Sellwood 1998, 2000).

\section{Bar Dynamical Friction in $N$-Body Simulations}
Fig. 1 plots the evolution of $\lag$ and $\len$ for a near-maximum disk,
and demonstrates that fast bars can survive for a Hubble time in such systems.
This simulation used a hybrid grid code, allowing us to achieve 
much higher resolution than in previous work.  At the end of the simulation, 
which lasted some 44 initial rotation periods of the bar, $\vpd = 1.3 \pm 0.1$.
Although the bar slows down by $\sim 35\%$ in this case, secondary bar growth
can still preserve a fast bar in this case.  But secondary bar growth has its 
limitations: for a flat rotation curve, ${\lag}_1/{\lag}_2 = {\om}_2/{\om}_1$, 
so a strongly braked bar would need to grow ever longer, which, however, 
makes friction even stronger (Debattista \& Sellwood 2000).

Weinberg \& Katz (2001) argued that $N$-body simulations require 
$N_{\rm p} \gtsim 4\times 10^6$ particles to resolve correctly
the bar-halo interaction, while simulations at lower $N_{\rm p}$ are largely 
noise-driven.  In Fig. 2 I plot the fractional drop in $\om$ between
early and late times in a series of simulations in which only $N_{\rm p}$ was 
varied.  No significant change in the evolution of $\om$ is seen, despite the 
factor of 100 difference in $N_{\rm p}$, suggesting that Weinberg \& Katz 
(2001) have been too pessimistic in their assessment of noise effects.  This 
can be understood by recognizing that a bar slows continuously, so that 
resonances are broadened.  If resonances had been sharp, then indeed large 
$N_{\rm p}$'s would be required, but for broad resonances, more modest 
$N_{\rm p}$'s suffice.

\begin{figure}[ht]
\plotone{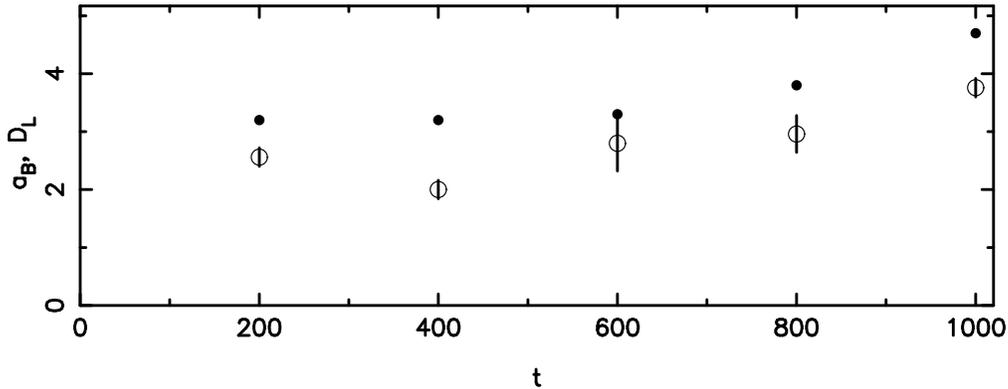}
\caption{Evolution of $\len$ (open circles with error bars) and $\lag$
(filled circles) for a massive disk simulation.  At $t=1000$, $\vpd = 1.3 
\pm 0.1$.}
\end{figure}

\begin{figure}[ht]
\plotone{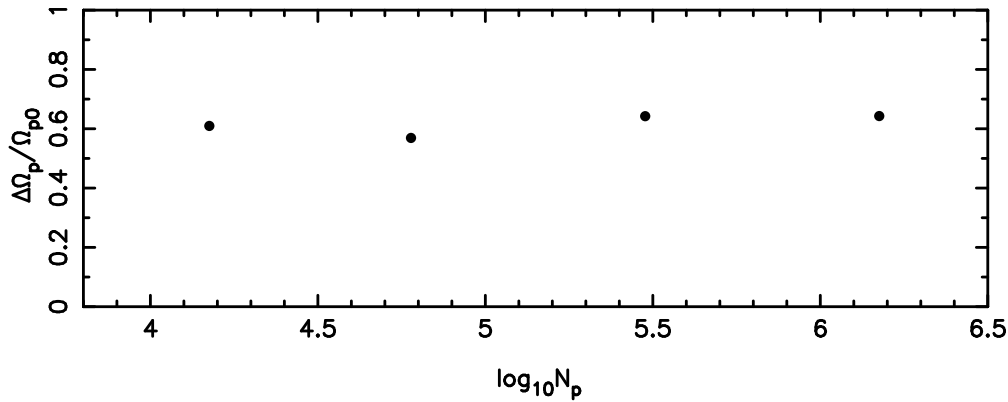}
\caption{The change in $\om$, due to halo dynamical friction, between 
early and late times for a series of simulations in which $N_p$ has 
been varied by a factor of 100.}
\end{figure}

\section{A Post-Interaction Galaxy}
Following Ostriker \& Peebles (1973), it is sometimes thought that 
unbarred (SA) galaxies must be stabilized by massive dark matter (DM) halos; 
however, bars can also be suppressed by rapidly rising rotation curves
(Toomre 1981; Sellwood \& Evans 2001).  Noguchi (1987) showed that disks 
stabilized by massive DM halos can still form bars via interactions, and
that such bars would be slow.  Thus SB galaxies with evidence of past 
interactions provide a means to statistically test whether SA galaxies are 
stabilized by massive halos.

NGC 1023 shows evidence of a past interaction (Sancisi et al. 1984), without 
being at present significantly perturbed.  Debattista et al. (2002) measured 
$\om$ for this galaxy, using the method of Tremaine \& Weinberg (1984).  The 
Tremaine-Weinberg (TW) method is contained in the following simple equation:
\begin{eqnarray}
\om \sin i = \frac 
{\int_{-\infty}^{\infty} h(Y) V_{los}(X,Y) \Sigma(X,Y)~dX~dY }
{\int_{-\infty}^{\infty} h(Y) X \Sigma(X,Y)~dX~dY} \equiv
\frac{\kin}{\pin}
\end{eqnarray}
where $V_{\rm los}$ is the line-of-sight velocity, $\Sigma$ is the surface 
brightness, $h(Y)$ is an arbitrary weight function, $i$ is the inclination 
and $(X,Y)$ are galaxy-centered coordinates along the disk's major and minor 
axes respectively.  We obtained 3 slit spectra parallel to the disk's major 
axis, for each of which we computed $\kin$ and $\pin$.  Then, plotting 
$\kin$ versus $\pin$ gives a straight line with slope 
$\om \sin i = 4.7 \pm 1.7$ \kmsa.  
From the rotation curve, corrected for asymmetric drift, this gives 
$\lag = {53^{+30}_{-15}} \arcsec$.  Multi-band photometry gives 
$\len = 69\arcsec \pm 5\arcsec$; therefore $\vpd = 0.77^{+0.43}_{-0.25}$, 
consistent with a fast bar.  Thus NGC 1023 must have a maximum disk; moreover, 
if the bar formed in the interaction (which cannot be ascertained), then it 
cannot have been stabilized by a massive DM halo.

\section{Pattern Speed in the Milky Way}
As shown by Kuijken \& Tremaine (1994), the TW method can also be applied
to the Milky Way Galaxy (MWG).   For discrete tracers, the TW method 
becomes:
\begin{eqnarray}  
\DV & \equiv & \om R_0 - V_{\rm LSR} \equiv \ 
    \frac{\mkin}{\mpin} - u_{\rm LSR} \frac{\msmt}{\mpin} \nonumber \\
    & = & \frac{ \sum_{i} f(r_i)v_{r,i}}{\sum_{i} f(r_i)\sin l_i 
\cos b_i} - u_{LSR} 
\frac{ \sum_{i} f(r_i)\cos l_i \cos b_i}{\sum_{i} f(r_i)\sin l_i \cos b_i} 
\end{eqnarray}
where $R_0$ is the Sun-MWG center distance, $V_{\rm LSR}$ is the tangential
velocity of the local standard of rest (LSR), $u_{\rm LSR}$ is the radial
velocity of the LSR, $f(r_i)$ is the observational detection probability 
(which need not be known), $v_{r,i}$ is the heliocentric radial velocity of 
a discrete tracer, and $(l_i,b_i)$ are its Galactic coordinates.  A survey
which satisfies the condition of uniform sampling, as required by the
method, is the ATCA/VLA OH 1612 MHz survey (Sevenster et al. 1997a,b \& 
2001), covering $|l| \leq 45\degrees$ and $|b| \leq 3\degrees$.  The
properties of OH/IR stars make them well suited for Galactic studies.  
They have bright, maser emission at 1612.23 MHz, which is insensitive to 
interstellar extinction; the double-peaked profile of the emission permits 
easy identification; and they are concentrated in the inner MWG, as a result 
of the Galactic metallicity gradient.
\begin{figure}[ht]
\plotone{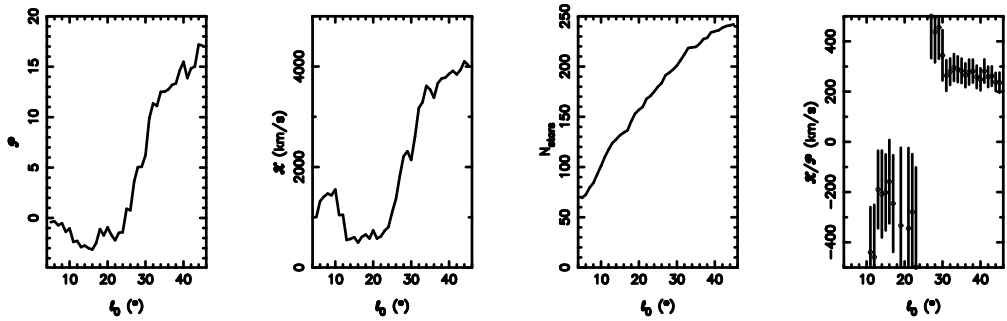}
\caption{The TW analysis of the OH/IR stars for changing $l_0$ (the
maximum $|l|$ in Eqn. 2).  From left to right are $\mpin$, $\mkin$, the 
number of OH/IR stars and the resulting $\mkin/\mpin$}
\end{figure}

We extracted from the ATCA/VLA OH 1612 MHz survey a sample of $\sim 250$ 
OH/IR stars which are older than 0.8 Gyr and have a flux density higher 
than 0.16 Jy; these criteria give OH/IR stars between $\sim 4$ and 
$\sim 10$ kpc away from the Sun.  Applying Eqn. 2 to this sample, as 
shown in Fig. 3, we obtained $\DV = 252 \pm 41$ \kms, where the error was 
estimated by resampling.  If we then assume $V_{\rm LSR}/R_0 = 220/8$ \kmsk\ 
(from SgrA$^*$ motion, Backer et al. 1999; Reid et al. 1999 and Cepheid 
proper motions, Feast \& Whitelock 1997) and $u_{\rm LSR} = 0$ (from SgrA$^*$ 
HI absorption spectrum, Radhakrishnan et al. 1980), we obtain 
$\om = 59 \pm 5 \pm 10$ \kmsk, where the last error is our estimated
systematic error.  This $\om$ is consistent with the hydrodynamical 
simulations of Bissantz et al. (2002), which also require 
$\vpd = 1.0 \pm 0.1$.

\end{document}